\global\newcount\refno
\global\refno=1 \newwrite\reffile
\newwrite\refmac
\newlinechar=`\^^J{} 
\def\refs#1#2{\the\refno\nref#1{#2}}
\def\nref#1#2{\xdef#1{\the\refno}
\ifnum\refno=1\immediate\openout\reffile=refs.tmp\fi %
             \immediate
\write\reffile{
 \noexpand\item{\noexpand#1. }#2\noexpand\nobreak} 
 \immediate\write\refmac{\def\noexpand#1{\the\refno}}
 \global\advance\refno by1}
\def\semi{;\hfil\noexpand\break ^^J} \def\nl{\hfil
\noexpand\break ^^J}
\def\refn#1#2{\nref#1{#2}} 

\def\rb#1{\raisebox{.8ex}{\scriptsize{#1}}} 
 
\refn\jacki{R.\ Jackiw, %
    {\noexpand\it Rev.\ Mod.\ Phys.\ } %
    {\noexpand\bf 52}, 661 (1980).} 
\refn\wuzee{Y.-S.\ Wu, A.\ Zee, %
    {\noexpand\it Nucl.\ Phys.\ } %
    {\noexpand\bf B258}, 157 (1985).} 
\refn\alvar{L.\ Alvarez-Gaum\'{e}, P.\ Ginsparg, %
    {\noexpand\it Nucl.\ Phys.\ B} %
    {\noexpand\bf 243}, 449 (1984).} 
\refn\mainz{A.\ Heil, A.\ Kersch, N.\ Papadopoulos, %
    B.\ Reifenh\"{a}user, F.\ Scheck, %
    {\noexpand\it J.\ Geom.\ Phys.\ } %
    {\noexpand\bf 7}, 489 (1990).} 
\refn\mcmua{D.\ McMullan, I.\ Tsutsui, %
    {\noexpand\it Ann.\ Phys.\ (N.Y.)} %
    {\noexpand\bf 237}, 269 (1995).} 
\refn\tanim{T.\ Tanimura, I.\ Tsutsui, %
    {\noexpand\it Ann.\ Phys.\ (N.Y.)} %
    {\noexpand\bf 258}, 137 (1997).} 
\refn\macke{G.\ W.\ Mackey, {\noexpand\it Induced %
    Representation of Groups and Quantum Mechanics} %
    (Benjamin, New York, 1969).} 
\refn\laidl{M.\ G.\ G.\ Laidlaw, C.\ M.\ DeWitt, %
    {\noexpand\it Phys.\ Rev.\ } %
    {\noexpand\bf D3}, 1375 (1971).} 
\refn\isham{C.\ J.\ Isham, in %
    {\noexpand\it Relativity, %
    Groups and Topology II} eds.\ B.\ S.\ DeWitt %
    and R.\ Stora (North-Holland, Amsterdam, 1984).} 
\refn\lands{N.\ P.\ Landsman, N.\ Linden, %
    {\noexpand\it Nucl.\ Phys.\ B } %
    {\noexpand\bf 365}, 121 (1991).} 
\refn\robso{M.\ A.\ Robson, %
    {\noexpand\it J.\ Geom.\ Phys.\ } %
    {\noexpand\bf 19}, 207 (1996).} 
\refn\dirac{P.\ A.\ M.\ Dirac, {\noexpand\it Lectures %
    on Quantum Mechanics} (Yeshiva, New York, 1964).} 
\refn\guill{V.\ Guillemin, S.\ Sternberg, %
    {\noexpand\it Symplectic techniques in physics} %
    (Cambridge University Press, Cambridge, 1984).} 
\refn\mcmub{D.\ McMullan, I.\ Tsutsui, %
    {\noexpand\it Phys.\ Lett.\ B.\ } %
    {\noexpand\bf 320}, 287 (1994).} 
\refn\armsa{J.\ M.\ Arms, %
    {\noexpand\it Math.\ Proc.\ Camb.\ Phil.\ Soc.\ } %
    {\noexpand\bf 90}, 361 (1981).} 
\refn\armsb{J.\ M.\ Arms, %
    {\noexpand\it Acta Phys.\ Polon.\ } %
    {\noexpand\bf B17}, 499 (1986).} 
\refn\mars{J.\ E.\ Marsden, %
    {\noexpand\it Lectures on Mechanics} (Cambridge %
    Univ.\ Press, Cambridge, 1992).} 
\refn\mont{R.\ Montgomery, %
    {\noexpand\it Lett.\ Math.\ Phys.\ } %
    {\noexpand\bf 8}, 59 (1984).} 
\refn\abra{R.\ Abraham, J.\ E.\ Marsden, %
    {\noexpand\it Foundations of Mechanics} %
    (Addison-Wesley, New York, 1978), 2nd edition.} 
\refn\babel{O.\ Babelon, C.\ M.\ Viallet, %
    {\noexpand\it Commun.\ Math.\ Phys.\ } %
    {\noexpand\bf 81}, 515 (1981).} 
\refn\naras{M.\ S.\ Narasimhan, T.\ R.\ Ramadas, %
    {\noexpand\it Commun.\ Math.\ Phys.\ } %
    {\noexpand\bf 67}, 121 (1979).} 
\refn\mitte{P.\ K.\ Mitter, C.\ M.\ Viallet, %
    {\noexpand\it Commun.\ Math.\ Phys.\ } %
    {\noexpand\bf 79}, 457 (1981).} 
\refn\danie{M.\ Daniel, C.\ M.\ Viallet, %
    {\noexpand\it Rev.\ Mod.\ Phys.\ } %
    {\noexpand\bf 52}, 175 (1980).} 
\refn\viall{C.-M.\ Viallet, in %
    {\noexpand\it Physics, Geometry, and Topology}, %
    eds.\ H.\ C.\ Lee (Plenum Press, New York, 1990).} 
\refn\singe{I.\ M.\ Singer, %
    {\noexpand\it Comm.\ Math.\ Phys.\ } %
    {\noexpand\bf 60}, 7 (1978).}  

\documentstyle[12pt]{article}
\begin{document}
\title{Inequivalent Quantizations of Gauge Theories}
\author{Kenichi Horie%
\thanks{Present address: High Energy Accelerator Research
Organization (KEK), Tanashi Branch, Tokyo 188-8501, Japan}\\ 
{\em Institut f\"{u}r Physik der 
     Johannes--Gutenberg--Universit\"{a}t,}\\ 
{\em  55099 Mainz, Germany} } 
\maketitle

\begin{abstract} 
It is known that the quantization of a system defined 
on a topologically non-trivial configuration space is 
ambiguous in that many inequivalent quantum systems are 
possible. This is the case for multiply connected 
spaces as well as for coset spaces. 
Recently, a new framework for these inequivalent 
quantizations approach has been proposed by McMullan 
and Tsutsui, which is based on a generalized Dirac 
approach. We employ this framework for the quantization 
of the Yang-Mills theory in the simplest fashion. The 
resulting inequivalent quantum systems are labelled by  
quantized non-dynamical topological charges. 
\vspace{2mm}\\
Number of pages: 24 
\end{abstract}
\newpage
%
           \section{Introduction}
In gauge theory, the highly non-trivial topology of the 
underlying gauge orbit space directly leads to or is 
related to many physically interesting phenomena. 
Some of the major examples are the $\theta$-vacua, 
topological terms like Chern-Simons term or Pontrjagin 
term, and also chiral anomaly, see e.g.\ Refs.\ %
\jacki, \wuzee, \alvar, and \mainz. 
The important point to note is that these facets are not 
consequences  of the non-trivial topology alone but arise 
only upon quantization or,  to be more distinct, upon 
reducing the simple quantum system on the total gauge 
potential space, which is affine, to the complicated 
gauge orbit space by dividing out the gauge group. 
Thus, in view of the prominent status of the gauge 
theories in physics, the quantization and reduction 
procedures of these systems must be investigated with 
ever more effort.   

In this short note we make first steps in applying the 
framework of  inequivalent quantizations proposed 
recently\rb{\mcmua,\tanim} to gauge theories and hint at 
further possible extensions. 

Many papers have appeared dealing with the methods of 
quantization on a non-trivial (finite dimensional) 
configuration space other than the well known case of 
linear space, see e.g.\ Refs.\ \mcmua, \macke, \laidl, 
\isham, \lands, and \robso\ and references therein as 
well as numerous other contributions. 
These works are of physical interest in their own 
rights, but the generalized Dirac approach of McMullan 
and Tsutsui\rb{\mcmua} based on the work of 
Mackey\rb{\macke} is directly applicable to gauge 
theory. To understand the ideas behind their approach, 
consider a classical system with first class 
constraints 
\begin{equation}\label{1}
    \phi_i = 0 
\end{equation} 
which, by definition, build up a closed algebra under 
the Poisson brackets. Due to Dirac\rb{\dirac} the 
quantization of such a constrained system may be 
performed by first quantizing the system on 
the entire phase space without the constraints and 
then imposing the constraints (\ref{1}) afterwards as 
operator conditions on quantum states, 
\begin{equation} 
    \hat{\phi_i} |\psi\rangle = 0 \;.
\end{equation} 
On the other hand, Mackey's approach to quantization 
on coset spaces takes a different path. The classical 
system is quantized directly  on the reduced space 
$G/H$ by generalizing the canonical commutation 
relations and using irreducible representations of 
what he called the imprimitivity relations.\rb{\macke} 
We will not go into details of this somewhat abstract 
algebraic procedure, but remark that Mackey's work was 
one of the first to show the existence of inequivalent 
quantum sectors on the coset space. In 
Ref.\ \mcmua\ this algebraic procedure was 
translated into the commonly used framework of 
Dirac formalism, thus, as we will see, 
making possible the application of inequivalent 
quantizations also to Yang-Mills theory. 
The generalization capable of incorporating the 
inequivalent quantizations within Dirac's approach 
consists in replacing the constraints (\ref{1}) by 
\begin{equation}\label{2} 
    \phi_i - K_i = 0 \;. 
\end{equation} 
The constraints (\ref{2}) define different classical 
systems which in case of coset space $G/H$ correspond 
to different foliations of the Marsden-Weinstein 
reduction.\rb{\guill} However, upon quantization the 
arbitrary numbers $K_i$ turn out to be discrete 
multiples of $\hbar$ and label the inequivalent 
quantum superselection sectors.\rb{\mcmub} 
This implies the important fact that in the 
classical limit $\hbar \longrightarrow 0$ all these 
inequivalent quantum systems reduce to the unique 
classical system described by (\ref{1}) and not to 
those classically different systems described by 
(\ref{2}). From standard viewpoint quantum physical 
systems play a more fundamental role in nature than 
the corresponding classical systems in that nature is 
described ultimatively by quantum rather than classical 
physics. Accordingly, regardless of actually how the 
quantum sectors have been derived, solely due to the 
fact that all these quantum theories have a unique 
classical limit, we may talk about `inequivalent 
quantizations' of this particular classical system, 
even though these quantum sectors have been obtained 
by quantizing different classical systems. These 
inequivalent quantum sectors are all legitimate 
physical systems, whereas among the classically 
inequivalent systems there is one distinguished 
system (\ref{1}) to which all quantum systems 
reduce to in the classical limit. 
The constraints (\ref{2}) are now partially second 
class (`anomalous') and care must be taken when 
quantizing. The resulting quantum theories are equipped 
with an induced connection\rb{\lands} and a generalized 
spin, i.e.\ the equations of motion on the coset space 
describe a particle with certain spin degrees of 
freedom depending on $K_i$ and minimally coupled 
to an external Yang-Mills type connection. 

One of the most prominent constrained physical system 
is gauge theory, where the true configuration space is 
obtained as the quotient of the affine gauge potential 
space by the gauge group. This structure is similar 
to the coset space structure considered in 
Refs.\ \mcmua\ and \mcmub, and it is 
interesting to know something about the possible 
inequivalent quantum systems which in the classical 
limit reduce to the given classical gauge theory. One 
of the questions we address in this paper is the 
physical role of the possible parameter which labels 
the quantum sectors. 

When this method of inequivalent quantizations is 
applied to Yang-Mills theory, the new constraints 
(\ref{2}) turn out to correspond to modified Gauss 
laws in general, see in this context also 
Refs.\ \armsa\ and \armsb. 
However, contrary to the case of homogeneous spaces, 
a number of legitimate modifications are possible, 
which are in general field dependent as anticipated 
in Ref.\ \mcmua. Among the possible modifications 
we pick up the most ``simple'' one. As a result, the 
modified Gauss laws induce non-dynamical charges. Upon 
quantization of the Yang-Mills theory these charges 
become quantized and describe topologically distinct 
field configurations. 

The organization of this paper is as follows. In order 
to apply these ideas to Yang-Mills theory we must be 
aware that the relevant configuration space is no 
longer a coset of the type $G/H$ but the base 
space ${\cal C}/{\cal G}$ of a principal bundle, 
where $\cal{C}$ is the total space of connections, 
and the structure group is given by the gauge group 
$\cal{G}$. (More precisely, we must restrict 
$\cal{C}$ to the dense set of irreducible connections 
since otherwise the group action is non-free.) Thus, 
as briefly mentioned in Ref.\ \tanim, the 
generalized Dirac quantization must be set up for 
principal bundles, which is the main task of the next 
section. In the third section, the geometric structures 
on the gauge orbit space are clarified and compared to 
the finite dimensional case, upon which the actual 
quantization is carried out. The last section discusses 
the results and hints at some further developments. 

    \section{Particle on a Principal Bundle}\label{s2}
In this section we extract from Refs.\ \mcmua\ %
and \tanim\ the mathematical structures essential 
for the inequivalent quantizations of Yang-Mills system 
and, in order to adapt the geometrical setting, consider 
a classical point particle moving on a principal bundle. 

Although such a physical system is already known in 
literature,\rb{\robso,\guill,\mars,\mont} the following 
presentation differs from what have been appeared 
hitherto in that it is kept as simple as possible and, 
furthermore, will allow a direct conversion to the 
Yang-Mills system in the next section. 

Let $G$ be a semisimple Lie group, such that its Lie 
algebra, denoted by ${\cal L}(G)$, possesses a 
non-degenerate metric ${\rm Tr}(RS)$, 
$R,S \in {\cal L}(G)$, defined by some multiple of 
the trace function in a certain irreducible matrix 
representation. Let $P$ be a principal bundle with 
this structure group $G$ over a basemanifold $B$ and 
consider the following $G$-invariant Lagrangian
\begin{equation}\label{lagrangian}
    L = \frac{1}{2}g_{\mu\nu}{\dot u}^\mu{\dot u}^\nu 
       - V(u)
      = \frac{1}{2}g(\dot{u},\dot{u})-V(u) \;, 
\end{equation}
where $\dot{u} \in T_uP$ is the velocity of the 
particle, and $V(u)$ and $g$ are $G$-invariant potential 
and metric, respectively. Thus, $V(u) = V(u\Lambda)$ and 
${R_{\Lambda}}_\ast g = g$, where $\Lambda \in G$ and 
${R_{\Lambda}}_\ast$ denotes the pull-back defined by 
the right group action $R_{\Lambda}(u)=u\Lambda$. The 
Hamiltonian corresponding to (\ref{lagrangian}) reads 
\begin{equation}\label{hamiltonian}
    H = \frac{1}{2}g(p,p)+V(u)\;,
\end{equation}
where $p$ is the particle momentum and $g$ in this 
equation stands for the inverse metric on the cotangent 
space $T^\ast_uP$. In what follows we identify the 
cotangent space with tangent space via the metric, and 
the mechanical system of the point particle has as its 
phase space the tangent bundle $TP$. This technical 
issue is not really essential but is introduced here 
to make the application of inequivalent quantizations 
to Yang-Mills theory in the next section more 
transparent. 

Having fixed the phase space and the Hamiltonian we now 
define the symplectic structure on the tangent bundle 
$TP$. For this consider on the cotangent bundle the 
canonical 1-form $\Theta_0 = p_\mu du^\mu$ and the 
canonical symplectic structure $\Omega_0 := -d\Theta_0$ 
derived from it. We pull back this symplectic 2-form 
$\Omega_0$ via the metric to the tangent bundle. Thus 
our symplectic 2-form reads $\Omega = -d\Theta$, where 
$\Theta$ is the pull-back of $\Theta_0$ given by  
\begin{equation}\label{symplectic} 
    \Theta = g_{\mu\nu}p^\mu du^\nu \;. 
\end{equation} 
Note that in this equation and in (\ref{hamiltonian}) 
the momentum is defined on the tangent space, 
$p=p^\mu\partial_\mu$. 
                                      
The $G$-symmetry of the system under consideration is 
expressed by the momentum maps $J$, see 
e.g.\ Ref.\ \abra. These are functions on the 
phase space $TP$ defined for each element 
$R \in {\cal L}(G)$ of the Lie algebra by 
\begin{equation}\label{momentum} 
    J(R)(p) := g(p,R^+) \;,  
\end{equation} 
where $R^+$ denotes the fundamental vector field of $R$. 
These momenta are conserved under the Hamiltonian flow. 
Furthermore, for the special symplectic structure 
$\Omega$ at hand, $J$ is a homomorphism of the Lie 
algebra of $G$ to that of functions on $TP$ under 
the Poisson bracket,\rb{\abra} 
\begin{equation}\label{poisson} 
    \{J(R),J(S)\} = J([R,S]) \;. 
\end{equation} 

Just as in the case of the symplectic structure 
$\Omega$, the momentum map $J$ (\ref{momentum}) is 
the pull-back via the metric of a corresponding 
canonical momentum mapping\rb{\abra} on the cotangent 
bundle. Thus both $\Omega$ and $J$ depend on the 
$G$-invariant metric $g$. Now the metric $g$ also 
defines another geometric structure on $TP$, 
namely a $G$-connection. Let us note here the link 
between this derived connection and the momentum map. 
The connection is given by the decomposition of the 
tangent space $T_uP$ into the vertical component 
$V_uP$ along the fibres and the orthogonal horizontal 
complement $H_uP$. If $\{T_a\}$ denotes an orthonormal 
basis of ${\cal L}(G)$, then $V_uP$ is spanned by 
${T_a}^+$, and the corresponding orthogonal projection 
of $T_uP$ onto $V_uP$ reads 
\begin{eqnarray}  
    \Pi : \; T_uP &\longrightarrow& V_uP \nonumber \\ 
    \phantom{\Pi : \;} 
              p   &\longmapsto& 
              g_u(p,{T_a}^+)\eta^{ab}{T_b}^+ 
    \label{pr} \;. 
\end{eqnarray} 
The matrix $\eta^{ab}$ is the inverse of 
$\eta_{ab}:=g_u({T_a}^+,{T_b}^+)$ and is generally 
$u$-dependent. It is easy to show that $\Pi$ is 
independent of the basis chosen. Now it is not 
difficult to confirm that the ${\cal L}(G)$-valued 
connection 1-form $\omega$ is obtained by simply 
replacing the fundamental vector field ${T_b}^+$ in 
(\ref{pr}) by its generator $T_b$, 
\begin{equation}\label{omega} 
    \omega(p) = g(p,{T_a}^+)\eta^{ab}T_b \;. 
\end{equation} 
Thus we see that the momentum to each generator $T_a$, 
\begin{equation}\label{ra} 
    r_a := J(T_a) \;, 
\end{equation} 
is linked to the connection via 
\begin{equation}\label{rw} 
    r_a(p) = {\rm Tr}(\omega(p)T_b)\eta_{ba} \;. 
\end{equation} 

Turning back to mechanics, to recover the dynamics of 
the particle on the base $B$, we will employ the Dirac 
constraints\rb{\dirac} in accordance 
with Ref.\ \mcmua. If we introduce constraints 
\begin{equation}\label{con_com} 
    r_a = 0 \;, 
\end{equation} 
then from (\ref{poisson}) we have 
\begin{equation}\label{con_algebra}
    \{r_a,r_b\} = f_{ab}{}^{\! c}\, r_c 
\end{equation}
with structure constants $f_{ab}{}^{\! c}$. Thus the 
$r_a$ build up a system of first class constraints 
consistent with the Hamiltonian, $\{r_a,H\}=0$, since 
momentum maps are conserved quantities. 

However, and this is the salient point recovered in 
Ref.\ \mcmua\ from the analysis of Mackey's 
inequivalent quantizations procedure, there is no 
stringent reason to restrict ourselves to the 
constraints (\ref{con_com}) but we are free to 
introduce (at the first place) an arbitrary element 
$K$ in the Lie algebra and replace (\ref{con_com}) by 
\begin{equation}\label{con_ine}
    r_a - {\rm Tr}(T_aK) = r_a - K_a = 0 \;, 
\end{equation}
which is formally the same as the modified constraints 
employed by \mbox{McMullan} and Tsutsui for the 
homogeneous space. The details of the physical 
motivation which led to this modified 
constraints may be found in Ref.\ \mcmua. 
Note that the reductions of the physical system via the 
modified constraints (\ref{con_ine}) corrospond to 
Marsden-Weinstein 
reductions.\rb{\guill,\robso,\mars,\mont} 
 
Let us remark that on a general principal bundle the 
modified constraints (\ref{con_ine}) are not the only 
possibility. In fact any modification will do, provided 
it reduces to the one considered in Ref.\ \mcmua\ %
for the homogeneous space. This restriction applies as 
long as we base our considerations on Mackey's work. 
Thus the full content of the idea of inequivalent 
quantizations is, insofar as its classical reduction 
part is considered, more general than the 
Marsden-Weinstein reduction considered above. 
Another natural modification of (\ref{con_com}) 
matching this condition is ($p \in T_uP$) 
\begin{equation}\label{con_ine 3}
    r_a(p) - g({T_a}^+,K^+) = J(T_a)(p - K^+) = 0 \;. 
\end{equation} 
The reason why we disregard this possibility is that 
the $K$-term is base point dependent in general and 
thus leads to a complication of the constraint algebra. 
                        
We shall now turn to the quantization of the so 
modified constrained system. The following discussion 
of quantization has been presented in 
Refs.\ \mcmua\ and \tanim\ in great 
detail and is therefore kept brief.

The new constraints (\ref{con_ine}) are no more all 
first class, since the Poisson bracket of two of them 
can not be expressed with these constraints alone but 
acquires an additional term $\mbox{Tr}\,[T_a,T_b]K$. 
This term vanishes if one of both generators lies 
in the kernel of the adjoint map $\mbox{ad}_K$. Thus 
(\ref{con_ine}) contains a first class subset given by 
\begin{equation}\label{con_first}
    \phi_s = J(T_s)-{\rm Tr}T_sK = r_s - K_s = 0 \;, 
\end{equation}
where $T_s$ are chosen to span 
${\rm Ker}(\mbox{ad}_K) =: s_K$, which decomposes as 
$s_K = t \oplus c$, where $t$ is the Cartan subalgebra 
containing $K$ and $c$ a possible orthogonal complement 
if $K$ is not regular semisimple. Let $S_K$ denote the 
Lie subgroup of $G$ generated by $s_K$. 

Constraints other than (\ref{con_first}) are second 
class, and for the constrained system there remains 
a gauge symmetry with respect to $S_K$-transformations. 
For these we introduce gauge fixing conditions 
by the functions 
\begin{equation}\label{gauge_fixing} 
    \xi_s = \xi_s(u,p) = 0 \;. 
\end{equation} 
The new system will be quantized by the path-integral 
method. In order to path integrate the Hamiltonian 
(\ref{hamiltonian}) over the phase space, we have to 
take into account both the first class constraints 
(\ref{con_first}), the other second class constraints 
from (\ref{con_ine}), and also the above gauge fixings. 
Denoting all these different constraints by $\phi_k$ 
we obtain 
\begin{eqnarray} 
    Z &=&\!\int \prod{\Omega_0}^{\! N}\, 
           \delta (\phi_k)\,{\rm det}^{\frac{1}{2}} 
           |\{\phi_k,\phi_{k'}\}| 
           {\rm exp}\left(\frac{i}{\hbar} 
           (\int \Theta_0 - H dt)\right) 
    \label{pi1}\\ %
      &=&\!\int\!{\cal D}p^\mu{\cal D}u^\nu 
           (\prod\det g) \,\delta (\phi_k)\, 
           {\rm det}^{\frac{1}{2}} 
           |\{\phi_k,\phi_{k'}\}| 
           {\rm exp}\left(\frac{i}{\hbar} 
           (\int\! dt\, g(p,\dot{u}) - H) \right)\quad
    \label{PI1} 
\end{eqnarray}
The first path-integral (\ref{pi1}) has been written on 
the cotangent bundle $T^\ast P$ using the product of 
the canonical Liouville form ${\Omega_0}^{\! N}$ over 
time $t$ and the canonical 1-form $\Theta_0$. Note that 
the measure $\prod{\Omega_0}^{\! N}$ is usually written 
more simply as ${\cal D}p_\mu{\cal D}u^\nu$, where 
$p_\mu$ and $u^\nu$ are canonical variables. The second 
expression (\ref{PI1}) is defined on the tangent bundle 
$TP$, where the determinant of the metric comes in owing 
to the change of variable $p^\mu = g^{\mu\nu}p_\mu$. 
Note that $\Theta = g_{\mu\nu}p^\mu du^\nu = 
g_{\mu\nu}p^\mu\dot{u}^\nu dt$. As explained in 
Ref.\ \mcmua, by a certain choice of the form 
of the gauge fixing conditions (\ref{gauge_fixing}), 
it is possible to integrate out the momentum. In so 
doing one can actually implement the constraints 
(\ref{con_ine}).  Concerning this calculation we note 
briefly that it is done by decomposing the momentum 
$p$ and the velocity $\dot{u}$ via the projection 
(\ref{pr}) into the corresponding vertical and 
horizontal components, whereupon the vertical 
projection $\Pi(p)$ (\ref{pr}) of the momentum $p$ can 
be entirely replaced by an expression of $K$. Finally, 
the following path-integral over the configuration 
space is obtained 
\begin{equation}\label{PI2}
    Z = \int {\cal D}u\,\prod{\rm det}g\,
        \delta(\xi_s){\rm det}|\{\phi_s,\xi_{s'}\}|\, 
        {\rm exp}\left(\frac{i}{\hbar}\int 
        L_{\rm tot}\,dt\right) \;, 
\end{equation}
where the total Lagrangian is a sum of the original 
Lagrangian (\ref{lagrangian}) projected onto the base 
space $B$ and additional terms linear and quadratic 
in $K$, 
\begin{equation}\label{l+dl}
    L_{\rm tot} = \frac{1}{2}g(\dot{v},\dot{v}) - V(u)
                   + {\rm Tr}(K\omega(\dot{w})) 
                   - \frac{1}{2}K_a K_b\eta^{ab}  \;. 
\end{equation} 
In this expression $\dot{v}$ and $\dot{w}$ denote the 
horizontal resp.\ the vertical velocity components. 
Thus the original Lagrangian part consisting of the 
first two terms may be projected down to the base space 
$B$ yielding 
$L_B = \frac{1}{2}g_B(\dot{x},\dot{x}) - V(x)$, where 
the metric $g_B$ on $B$ is defined by the original 
metric $g$ acting on some horizontal lift of $\dot{x}$ 
on $B$. As for the last quadratic Casimir-expression 
(cf.\ Refs.\ \tanim\ and \lands\ it also 
is a function on the base $B$, since $\eta^{ab}$ is 
$G$-invariant. The $K$-linear term depending on the 
vertical fibre velocity $\dot{w}$ via the connection 
1-form $\omega$ (\ref{omega}) is gauge-dependent. 

Let us calculate the change of this additional term 
under an $S_K$-trans\-form\-ation. Let $u \mapsto us$ 
be an $S_K$-transformation, which can be written as 
$s=e^{\theta^r T_r}e^{\xi^p T_p}$ with the help of 
bases $\{T_r\}$ and $\{T_p\}$ of $t$ and $c$ in the 
orthogonal decomposition $s_K=t\oplus c$. Taking into 
account the defining properties of a connection, the 
gauge change in $L_{\rm tot}$ is calculated to be a 
total time derivative only, 
\begin{equation}\label{symmetry}
    \Delta L_{\rm tot} =
    \Delta L_K = \frac{d}{dt}\mbox{Tr}(K\theta) \;, 
\end{equation}
which implies the $S_K$-symmetry at the classical 
level. The quantum $S_K$-symmetry of the theory on 
the other hand requires the path-integral 
(\ref{PI2}) to be indepent of the gauge fixing. This 
is the case if and only if paths related to each other 
by a gauge transformation contribute to the 
path-integral with the same amplitude. This requirement 
is stronger than the classical one in (\ref{symmetry}) 
and leads to a quantization of $K$: consider a path 
$u(t)$, $t \in [0,T]$, in (\ref{PI2}). Gauge related
paths contributing to the transition from $u(0)$ to 
$u(T)$ are given by
\begin{equation}\label{2 paths} 
    u_s(t) = u(t) s(t)\,,\qquad s(t)\in S_K\,, 
    \qquad s(0) = s(T) = 1 \;.
\end{equation}
The amplitude of $u_s(t)$ differs from that of $u(t)$ 
by an phase given by
\begin{equation}\label{phase dif} 
    \Delta = \frac{1}{\hbar}\int_{0}^{T} 
             dt\,\Delta L_{\rm tot}
           = \frac{1}{\hbar}{\rm Tr}
             \left(K\left(\theta(T)-\theta(0) 
             \right)\right) \;. 
\end{equation}
The $S_K$-symmetry thus requires 
$\Delta \in 2\pi{\bf Z}$. This together with the 
periodicity property of $S_K$, 
$\theta^r(T)-\theta^r(0) \in 2\pi {\bf Z}$ implies 
that each component $K^r$ of $K$ must be quantized 
in multiples of $\hbar$, labelling the inequivalent 
quantizations.\rb{\mcmua} 

As mentioned already in the introduction, note that, 
since the components $K^r$ of $K$ are discrete 
multiples of $\hbar$,\rb{\mcmub} they all become 
zero in the classical limit. Thus, although the 
inequivalent quantum systems have been obtained by 
quantizing different classical systems, which 
correspond to different foliations of the 
Marsden-Weinstein reduction,\rb{\robso} according to 
our philosophy explained in the introduction we view 
these quantum systems as {\em inequivalent 
quantizations} of the same classical system to which 
all these quantum sectors reduce in the classical limit. 

Inequivalent quantizations considered so far are of 
physical interest when quantizing a classical system 
on a given topologically non-trivial configuration 
space $B$. One seeks for a principal bundle structure 
$P$ with a total space tractable for quantization. For 
example if the base space is not simply connected, one 
can try first a quantization on its universal covering. 
If this works, then, besides the ``canonical'' 
quantization on $B$, we automatically obtain a series 
of inequivalent quantum systems. These may be 
considered to be a result of the non-trivial 
topology of the base space.

The quantization method discussed so far however 
suffers from an arbitrariness of the geometry chosen 
on $P$, especially the form of the metric employed. 
Thus in order the above procedure to be of real 
interest the geometric structure on $P$ must be such 
that it arises naturally from the mathematical or 
physical setting. For example in case of a Lie group 
$G$ over the homogeneous space $G/H$ as considered in 
Refs.\ \mcmua\ and \tanim\ the metric on $G$ 
is given by an $\mbox{ad}_G$-invariant one. 
In what follows we shall consider gauge theories. 
In this case the base configuration space, which is 
the quotient of the affine space of connection forms 
and the gauge transformation group, has a complex 
topology in contrast to the affine total space, on 
which a natural metric can be chosen.

        \section{Yang-Mills system}\label{s3}
In this section we apply the inequivalent 
quantizations method to gauge theory. For simplicity 
we take the structure group $G$ to be $\mbox{SU}(N)$. 
As is well known the action of the gauge group makes 
the space of connections a principal 
bundle.\rb{\babel,\naras} More precisely, in this 
paper we consider gauge potentials $A_\mu$ on 
${\bf R}\times\Sigma$, where ${\bf R}$ denotes the 
time coordinate and $\Sigma$ is the 3-sphere. This 
topology of spacetime may arise e.g.\ if gauge fields 
defined on the Minkowski spacetime ${\bf R}^4$ fall off 
at space-like infinity fast enough so that they can be 
considered as fields on ${\bf R}\times\Sigma$. Also, in 
the following discussion more general cases where 
$\Sigma$ is an arbitrary closed 3-manifold may be 
treated analogously. Let $P$ be a principal $G$-bundle 
and ${\cal C} := \{A=A_i dx^i\, | i=1,2,3 \}$ the 
space of irreducible space-component connection 
1-forms. Let $\cal G$ denote the gauge group of all 
time-independent gauge transformations, which are 
sections into the adjoint group bundle defined 
by $P \times_{\rm Ad} G$. Then the action of 
$\cal G$ on $\cal C$ makes it a principal 
$\cal G$-bundle over the orbit space.\rb{\naras} The 
space $\cal C$ is an affine space and has a natural 
metric determined by the space components of the 
spacetime metric and the Killing metric of the 
structure group: Let ${\rm ad}P$ be the adjoint 
vector bundle defined by $P \times_{\rm ad}{\cal L}(G)$, 
where the structure group $G$ acts on its Lie algebra 
${\cal L}(G)$ by the adjoint representation. Then the 
canonical non-degenerate positive metric is defined on 
the space $\Gamma^m$ of ${\rm ad}P$-valued $m$-forms,  
$\Gamma^m = \Lambda^{\!m}(\Sigma)\otimes \mbox{ad}P$, by 
\begin{equation}\label{(a,b)}
    (\alpha , \beta) 
    = \int_{\Sigma}^{} 
      {\rm tr}(\alpha \wedge {}^\ast \beta) 
    = \int_{\Sigma}^{}\, 
      \frac{1}{m!}\alpha^a_{\mu_1\cdots\mu_m} 
      \beta_a^{\mu_1\cdots\mu_m}\sqrt{g}\,d^3\! x \;, 
\end{equation}
where $\ast$ denotes the Hodge operator and $g$ stands 
for the determinant of the metric. Since the tangent 
space at any point in $\cal C$ is the space 
$\Gamma^1$ of ${\rm ad}P$-valued 1-forms, 
eq.\ (\ref{(a,b)}) defines a canonical gauge-invariant 
metric on $\cal C$ and thus determines a well-defined 
connection on $\cal C$,\rb{\babel} see below. Note that 
$\Gamma^0$ of sections of ${\rm ad}P$ is the Lie algebra 
of the gauge group $\cal G$ (see e.g. 
Ref.\ \mitte), which we denote by $\cal L(G)$. 
It is equipped with the ad-invariant 
metric (\ref{(a,b)}). 

The covariant derivative 
$\nabla : \Gamma^m \rightarrow \Gamma^{m+1}$ and its 
adjoint 
$\nabla^\ast : \Gamma^m \rightarrow \Gamma^{m-1}$ with 
respect to the metric (\ref{(a,b)}) are given by
\begin{equation}
    \nabla \alpha = d\alpha + [A \,,\, \alpha] \;,\quad 
    (\nabla^\ast\alpha ,\beta) := (\alpha ,\nabla\beta) 
    \;,
\end{equation}
where $\alpha \in \Gamma^m$ and 
$\beta \in \Gamma^{m-1}$.

Let us now turn to the standard formulation of 
Yang-Mills theory itself.
The curvature and the action are given by
\begin{eqnarray}\label{ym-system}
    F_{\mu\nu} &=& 
    \partial_\mu A_\nu - \partial_\nu A_\mu 
                  + [A_\mu ,A_\nu]          \\ 
    I          &=& 
    \int dt L \; := \; 
    \int dt\int_{\Sigma}^{}(\sqrt{g}\,d^3\! x)\, 
    \frac{1}{4}{\rm tr}(F_{\mu\nu}F^{\mu\nu}) \;. 
\end{eqnarray}

In the framework of canonical 
formalism\rb{\jacki,\babel} we 
make use of gauge invariance and set $A_0 = 0$. The 
coordinate variables are the space components of the 
potential. Denoting the partial time derivative by 
a dot, the electric and magnetic field strength 
components read 
\begin{eqnarray}\label{E and B}
    E &=& F^0{}_{\! i}\,dx^i \; =\; \dot{A} \;, \\
    B &=& \frac{1}{2}\epsilon_{ijk}F^{jk}dx^i \;, 
\end{eqnarray}
where $A=A_i dx^i$. The canonical momenta are given by 
\begin{equation}\label{ym-momenta}
    \pi = \pi_i^a T_a dx^i
        = \frac{\delta L}{\delta\partial_t A^i_a} 
          \,T_a\, dx^i
        = \dot{A} \; , 
\end{equation}
and the Hamiltonian by 
\begin{equation}\label{ym-hamilton}
    H = \frac{1}{2}(\pi,\pi) + \frac{1}{2}(B,B) \;. 
\end{equation}
As in the case of a point particle on a principal 
bundle discussed in the previous section the momentum 
$\pi$ is defined on the tangent space. 
The supplementary Gauss conditions 
\begin{equation}\label{gauss}
    \nabla^\ast\pi = 0 \; 
\end{equation}
express the gauge invariance of the Yang-Mills theory 
and constrain it from the whole connection space 
$\cal C$ to the gauge orbit space. To show that the 
Gauss laws exactly correspond to the constraints 
(\ref{con_com})\rb{\armsa,\armsb} and moreover that 
the whole geometrical settings match those of the 
previous section let us have a closer look at the 
differential geometric structure of 
the $\cal G$-principal 
bundle $\cal C$.\rb{\babel,\danie,\mitte,\viall,\singe} 
Consider an infinitesimal gauge transformation 
$g = e^\theta$ of the potential 
\begin{equation}\label{gauge trans}
    A \mapsto A^g := g^{-1}Ag + g^{-1}dg
                  \approx  A  + \nabla\theta \;.
\end{equation}
This can be interpreted as follows: The expression 
$\nabla\theta$ is an element of $\Gamma^1$ and is 
the tangent vector in $T_A{\cal C}$ generating the 
gauge transformation on $\cal C$. In fact it is the 
fundamental vector field related to $\theta$, which 
is an element of $\Gamma^0$ equivalent to the Lie 
algebra $\cal L(G)$. Thus the image of the covariant 
derivative $\nabla$ in the tangent space at a point 
$A \in {\cal C}$ is given by the vertical tangent 
vectors along the gauge fibres of $\cal C$ considered 
as $\cal G$-principal bundle, which we denote 
by $V_{\! A}{\cal C}$. To construct its orthogonal 
complement, let $\Box := \nabla^\ast\nabla$ be the 
covariant Laplacian acting on $\Gamma^1$ and 
let ${\rm G} = \Box^{-1}$. To be precise, this inverse 
is defined if the connection is irreducible, the set of 
which however is known to be dense 
in $\cal C$.\rb{\singe} 
The vertical projection is then defined by
\begin{equation}\label{projection}
    \Pi := \nabla{\rm G}\nabla^\ast \;, 
\end{equation}
and the horizontal component of $T_A{\cal C}$ is given 
by $H_A{\cal C} := (1-\Pi)T_A{\cal C}$. The 
corresponding connection $\omega$ of the orthogonal 
decomposition 
$T_A{\cal C} = H_A{\cal C} \oplus V_{\! A}{\cal C}$ 
reads\rb{\viall}
\begin{eqnarray} 
    \omega :\; 
    T_A{\cal C}    & \longrightarrow & \; {\cal L(G)} 
    \nonumber \\ 
    \phantom{\omega :\;} 
    \tau\;\;       & \longmapsto     & %
    {\rm G}\nabla^\ast\tau \;. 
    \label{connection}
\end{eqnarray}
It is easy to check that $\omega$ so defined is 
indeed a connection. 

By analogy with (\ref{momentum}) the momentum 
functional reads ($\theta \in {\cal L(G)}$) 
\begin{eqnarray} 
    J(\theta)(\pi) &=& (\pi,\nabla\theta) 
                  \;=\;  (\nabla^\ast\pi,\theta)  
    \label{mom1} \\
                   &=& (\Box\omega(\pi),\theta) \;. 
    \label{mom2} 
\end{eqnarray} 
So the momentum is linked to the connection via $\Box$, 
which plays the role of what was called $\eta$ in 
(\ref{rw}). The constraints corresponding to 
(\ref{con_com}), when written without components, are 
exactly the Gauss laws (\ref{gauss}). In order to apply 
the method of inequivalent quantizations, we refer 
to the modifications of the constraints 
(\ref{con_ine}) in the last section and thus modify 
the Gauss laws as 
\begin{equation}\label{gauss-ine1}  
    \nabla^\ast\pi - K = 0 \;, 
\end{equation} 
with field-independent $K \in {\cal L(G)}$. As has 
been indicated by (\ref{con_ine 3}) for the finite 
dimensional case, also in Yang-Mills system there are 
more than one modification procedure possible. As an 
example, the one paralleling with (\ref{con_ine 3}) 
would read 
\begin{equation}\label{gauss-ine2} 
    \nabla^\ast\pi - \Box K = 0 \;. 
\end{equation} 
This constraint shift is field-dependent, and its 
Poisson algebra becomes rather untractable, contrary 
to that of (\ref{gauss-ine1}), which is straightforward. 
We therefore modify the Gauss constraints by means of  
(\ref{gauss-ine1}). This form of modification can be 
obtained also by simple guess, but we must hold in 
mind that it is a very special case among possible 
other modifications. 

Employing the standard canonical symplectic structure, 
we find that the Poisson commutation relation for these 
modified constraints reads now 
\begin{equation}\label{ym-com}
    \{(u,\nabla^\ast\pi - K) \,,\, %
      (v,\nabla^\ast\pi - K)\} 
    = \left([u,v],\nabla^\ast\pi - K \right)
    + \left([u,v],K \right) \;. 
\end{equation}
The last term in this equation spoiling the closedness 
of the algebra vanishes whenever $u$ or $v$ are 
sections in the kernel 
bundle $\mbox{Ker}(\mbox{ad}_K)$. For simplicity of 
the following discussions let us assume that $K(x,t)$ 
is a regular semisimple element for all 
$x \in \Sigma$ (so that the kernel bundle is indeed 
well-defined) and, furthermore, that the kernel bundle 
is trivial. We may then take normalized base vector 
fields $\sigma_s$, $s=1,...,\dim(s_K)$, and obtain the 
components of the first class constraints as
\begin{equation}\label{ym first}
    (\nabla^\ast\pi,\sigma_s) - (K,\sigma_s) = 0 
    \;\;\Leftrightarrow\;\;     
    (\nabla^\ast\pi)_s - K_s = 0 \;.
\end{equation}
Since $K$ does not depend on fields $A$ and $\pi$, all 
components of the constraints (\ref{gauss-ine1}), the 
first class as well as the second class ones, commute 
exactly with the Hamiltonian and thus, if $K$ is 
assumed to be time-independent, 
$\frac{\partial}{\partial t}K = 0$, all constraints 
are preserved in time. Since the extra term $([u,v],K)$ 
in (\ref{ym-com}) is field-independent, there are no 
additional second order constraints. Due to the presence 
of second class constraints in (\ref{gauss-ine1}) the 
full gauge symmetry is lost and there is only a 
residual $S_K$-symmetry.

We now consider the quantization of the modified 
Yang-Mills system at hand by means of path-integral 
method. As described in the previous section we 
introduce gauge fixings for the $S_K$-symmetry and 
include all the constraints into the phase space 
path-integral analogously to (\ref{PI1}). The delta 
functionals for the first class constraints 
(\ref{ym first}) are implemented in a standard way 
into the phase factor by using the time component of 
the connection $A_0$.\rb{\jacki} The background of the 
calculational steps in the path-integral based on 
the canonical theory of Yang-Mills system have been 
explained in detail in the review by Jackiw\rb{\jacki} 
and will not be mentioned here. 
Performing the $\pi$-integration one is finally led 
to a modified Lagrangian
\begin{eqnarray}\label{ym l+dl}
    L_{\rm tot} &=& L + L_K  \nonumber \\
                   &=& L + (A_0 ,K) \;.
\end{eqnarray} 
Contrary to the finite-dimensional case (\ref{l+dl}) 
the Casimir-term is missing in (\ref{ym l+dl}). This 
is due to the fact that the delta functionals have 
not been evaluated completely as in the previous 
section but only implemented into the Lagrangian by 
employing the time-component $A_0$ as Lagrange 
multiplier. Thus the $K$-quadratic term is implicitly 
contained in the Lagrangian, which becomes explicit 
when the field equation $\nabla^\ast E = K$ is taken 
into account. The time-component is used to maintain 
Lorentz covariance, see also below. 

Under an infinitesimal $S_K$-gauge transformation 
$g = e^\theta$, where $[\theta,K] = 0$, the Lagrangian 
changes by a total time derivative only, 
cf.\ eq.\ (\ref{symmetry})
\begin{equation}
    \Delta L_{\rm tot}
    = (\dot{\theta}+[A_0 ,\theta]\,,\, K)
    = \frac{d}{dt}(\theta,K) \;.
\end{equation} 

The quantization conditions of the component 
fields $K_s$ may be obtained exactly in the same 
way as described in (\ref{2 paths}) and 
(\ref{phase dif}) and yields
\begin{equation}\label{quant of K} 
    \frac{1}{\hbar}(\theta_s(T)-\theta_s(0)\,,\, K) = %
    \frac{2\pi n}{\hbar} \int_\Sigma^{}K_s    \in  
    2\pi{\bf Z} \;, 
    \quad n \in {\bf Z} \;, 
\end{equation} 
where $\theta_s(x,t)$, $t \in [0,T]$, is a 
time-dependent $S_K$-gauge transformation with 
the boundary conditions 
$e^{\theta_s(0)} \equiv e^{\theta_s(T)} \equiv 1$. In 
view of the equations of motion, $\nabla^\ast E = K$, 
we conclude that the charges associated with the Cartan 
subgroup of $G$ are quantized, 
\begin{equation}\label{charge quant} 
    Q_s := \int_\Sigma^{}(\nabla^\ast E)_s \in 
    \hbar\cdot{\bf Z} \;, 
\end{equation}
and, owing to the abelian nature of the residual Cartan 
subgroup, these external charges are covariantly 
conserved and static. 
As in the introduction and also in the previous section 
we note that, since these external charges $Q_s$ in 
(\ref{charge quant}) are multiples of $\hbar$, all the 
inequivalent quantum sectors of the Yang-Mills system 
reduce to the distinguished classical system described 
by the usual Gauss law (\ref{gauss}) in the classical 
limit $\hbar \longrightarrow 0$. 

In (\ref{ym l+dl}) it is not apparent that 
$L_K$ induces a topological charge. This is so because 
the Lorentz covariance was lost in applying the 
canonical formalism. However, we may recover the 
Lorentz covariance in the following way. We first 
look at the action of the $K$-linear Lagrangian 
part in the time interval $[0,T]$ under consideration, 
\begin{equation}\label{wk} 
    I_K = \int_{0}^{T}L_K dt 
        = \int_{[0,T]\times\Sigma}^{}A_0^a K_a dV \;, 
\end{equation} 
where $dV:=\sqrt{g}\,d^3\! x dt$ is the 4-volume form. 
Since the entire expression in the spacetime integral 
is Lorentz invariant and $A_0^a$ is the 0-component of 
a 4-form, $K_a$ itself must be the 0-component of some 
$K^\mu_a$, say. Also, the requirement of time 
independence of $K_a$ stated before has as its 
covariant counterpart the vanishing of the divergence 
$\partial_\mu(K^\mu_a\sqrt{g})$. Using the language 
of differential forms, we may express the integrand 
in (\ref{wk}) as $i_K A \wedge dV$, where $i_K$ denotes 
the interior product of forms and vectors. This 
expression is equivalent to $A \wedge i_K dV$, and 
the 3-form $\tilde{K}:=i_K dV$ so introduced is exact. 
The phase difference $\Delta W_K$ induced by the above 
mentioned $S_K$-gauge transformation may be derived, 
using $\partial\Sigma=\emptyset$, as 
\begin{eqnarray} 
   \Delta I_K &=& \int_{[0,T]\times\Sigma}^{}
                  {\rm tr}\nabla\theta\wedge\tilde{K} 
             \;=\;\int_{[0,T]\times\Sigma}^{} 
                  {\rm tr}\,d\theta\wedge\tilde{K} 
                  \nonumber\\ 
              &=& \int_{\partial([0,T]\times\Sigma)}^{} 
                  {\rm tr}\,\theta\tilde{K}        
                  \nonumber\\ 
              &=& \int_\Sigma^{}{\rm tr}
                  (\theta(T)-\theta(0))\tilde{K} \;. 
              \label{dwk} 
\end{eqnarray} 
The quantum condition (\ref{quant of K}) now reads 
\begin{equation}\label{top} 
    Q_s = \int_\Sigma^{}(\nabla^\ast E)_s 
        = \int_\Sigma^{}\tilde{K}_s \in 
          \hbar\cdot{\bf Z} \;, 
\end{equation} 
where the exact forms $\tilde{K}_s$ represent 
elements in the de Rham cohomology group, 
$[\tilde{K}_s] \in {\rm H}^3(\Sigma ;{\bf R})$. In 
this way the charges associated with the residual 
Cartan subgroup are linked to the topology of the 
underlying space $\Sigma$. If gauge-coupled fermionic 
matter is included, the whole quantization procedure 
goes through without changes, and on the lefthand 
side of (\ref{top}) matter charge adds up. Equation 
(\ref{top}) also states the absence of any 
superselection sector if the topology of the space 
is trivial, i.e.\ ${\rm H}^3(\Sigma ; {\bf R})=0$. 

             \section{Discussion and Outlook}
In the previous section we have employed the method of 
inequivalent quantizations in the simplest fashion, 
where a field-independent constraint shift was chosen. 
As a result the Yang-Mills quantum system contains 
superselection sectors labelled by quantized external 
charges arising from the non-trivial topology of the 
underlying space. 

These quantized charges have nothing to do with the 
topological charge (Pontryagin index) of instantons, 
since in our case the residual Cartan symmetry is 
abelian. Contrary to those topological charges, which 
directly give insight into the topology of the 
configuration space ${\cal C}/{\cal G}$ in that they 
label the connected components, the inequivalent 
quantizations induced by the simple modification 
(\ref{gauss-ine1}) of the Gauss constraints could 
only detect the topology of the underlying 
space $\Sigma$. However, the idea of inequivalent 
quantizations as explained especially in 
Ref.\ \mcmua\ can incorporate more general 
quantum constraints than those induced by the 
modified Gauss laws (\ref{gauss-ine1}) corresponding 
to Marsden-Weinstein reductions.\rb{\guill} With a 
more systematic analysis of other field-dependent 
modifications like (\ref{gauss-ine2}) it may be 
possible to unveil the quantum implications of 
the topology of configuration space itself. 
Especially, one should look for induced Lagrangian 
parts which are explicitly topological, since then 
we can be fairly sure that they are closely linked 
to the topology of the gauge orbit space, as can be 
seen from the work of Wu and Zee.\rb{\wuzee} 

Another remaining issue is the investigation of 
the induced connection\rb{\lands,\mcmua} on the gauge 
orbit space. For example, the one induced by the 
above choice (\ref{gauss-ine1}) is seen to be 
non-exact in the functional sense, and one might 
wonder whether it is possible to relate these induced 
connections to the Chern classes of the orbit space, 
which is known to be a classifying space for the 
gauge group $\cal G$. \vspace{8mm}\\ 
{\bf Acknowledgements: } 
I would like to greatly thank Prof.\ Izumi Tsutsui
for not only suggesting this line of research but also
for his many helpful discussions; moreover, I am very 
much indebted to Prof.\ Martin Kretzschmar, who helped
to resolve an organizational problem 
during the preparation of this work. 

\newpage
  \vfill\eject\immediate\closeout\reffile
  \centerline{{\large\bf References}}\bigskip%
  \frenchspacing\noindent\begin{description}%
  \input refs.tmp\end{description}\vfill\eject%
  \nonfrenchspacing 

\end{document}